\documentclass[intlimits,twoside,a4paper]{article}

\usepackage[active]{srcltx}

\usepackage{amsmath,amssymb}
\usepackage{graphicx}
\usepackage{wrapfig}
\usepackage{color}

\usepackage[T2A]{fontenc}
\usepackage[cp1251]{inputenc}
\usepackage{cmpj2}

\issue{2012}{15}{3}{33801}

\doinumber{10.5488/CMP.15.33801}

\title[Solvation forces between polyelectrolyte layers ]%
{Solvation force between tethered polyelectrolyte layers. A density functional approach}

\author[O. Pizio  \textsl{et al.}]{O. Pizio\refaddr{label1}, A. Patrykiejew\refaddr{label2},
S. Soko\l owski\refaddr{label2},
J.M. Ilnytskyi\refaddr{label3}}
\addresses{
\addr{label1} Instituto de Qu\'{i}mica, Universidad Nacional Autonoma de M\'{e}xico,
Circuito Exterior, Ciudad Universitaria, \\ 04360~M\'{e}xico, D.F., M\'{e}xico
\addr{label2} Department for the Modelling of Physico-Chemical Processes, Maria
Curie-Sk\l odowska University, \\ 20031~Lublin, Poland
\addr{label3}Institute for Condensed Matter Physics of the National Academy of Sciences of Ukraine,\\ 1~Svientsitskii Str., 79011~Lviv, Ukraine
}

\date{Received June 14, 2012}

\authorcopyright{O. Pizio, A. Patrykiejew, S. Soko\l owski, J.M. Ilnytskyi, 2012}

\begin{document}

\maketitle

\begin{abstract}
We use a version of the density functional theory to study the solvation
force between two plates modified with a tethered
layer of chains. The  chains are built of tangentially jointed charged spherical segments. The
plates are immersed in an electrolyte solution that involves
cations, anions and solvent molecules. The latter molecules are
modelled as hard spheres.  We study the dependence of the solvation
force and the structure of chains and of solute molecules on
the grafting density, length of chains,
architecture of the chains and on concentration of the solute.

\pacs 82.35.Rs, 61.25.H-, 82.35.Gh, 82.35.Rs
\end{abstract}

\section{Introduction}

Layers of charged polymers tethered onto solid
surfaces have attracted great attention for the recent years due to
their biological importance and numerous
practical applications~\cite{i2}.
The behavior of charged polymers at interfaces is not only more
complex as compared to charged polymers in bulk solution~\cite{i2a,i22a,i2b,i2d}
but is also
more complex than the behavior of systems involving
uncharged tethered polymers~\cite{iaa,ibb,icc}.
Generally, these systems contain polymers carrying
positive and/or negative charges, surface
charges, mobile ions in solution and solvent molecules.
All these components make the properties
of a system intricately dependent on electrostatic
interactions~\cite{i3,i5,i6,i7,i8}.

A special type of charged polymers are polyampholytes, i.e.,
polymers consisting of
positively and negatively charged segments and,
in some cases,
also neutral
segments.  Differently charged, as well as
neutral segments, can be distributed randomly
or in a certain sequence~\cite{i10,i10a,i10b,i10c}.
A number of synthetic
and natural (e.g., proteins) polymers can be classified as
polyampholytes~\cite{i13,i13a,i13b}
A special example of such systems are the so-called
polyzwitterionic polymers~\cite{i15}.

Due to their
unique features, polyampholytes have a wealth of important
and practical applications, e.g., as
dispersing additives, gelling
agents, rheology modifiers, etc.~\cite{i13}. Nevertheless,
polyampholytes have received much less
attention, particularly in theoretical studies, as compared to
neutral polymers.
Consequently, our understanding of polyampholytes, especially
tethered polyampholytes, is far from being complete.

In the case of uncharged solid surfaces and
tethered chains whose segments bear the same charges,
the electrostatic repulsion forces can be strong even at
low grafting densities. Consequently,
under such conditions, the chains
are stretched  and the system enters
the brush regime~\cite{i10c}.
 In the case of tethered
polyampholytes, the pinned layer behaves like a polyelectrolyte
when the chains are globally charged and the
resultant single chain charge is significant.
However, for zero or for low net charge,  the polyampholyte
can coil and the tethered layer collapses. Low density
tethered polyampholytes exhibit the highest deformation amplitude
among the stretched and the collapsed states~\cite{i18,i20}.
The structure and thermodynamic
properties of tethered polyelectrolytes have been investigated
 using both theoretical~\cite{i10b,i21,i23,i23b,i23c,i23d,i23e} and
computer simulation~\cite{i10a,i10c,i24,i25,i26,i29,i30,i34}  methods.
Density functional methods
are among the most successful theoretical approaches~\cite{d0,d1,d2,d3,d4,d5}.

Interaction forces between surfaces and colloidal stability are
closely related. A stable colloidal suspension is characterized by repulsive forces
between the colloidal particles, while attractive forces lead to an
unstable suspension. One might expect that pinned polyelectrolyte
layers induce repulsive forces and thus it leads to suspension
stabilization. However,
it has been also
observed that polyelectrolytes
may destabilize colloidal suspensions~\cite{e2}.
The problem of evaluation of the solvation forces between
the surfaces modified with tethered charged chains
and their dependence on the properties of a solvent and
on the properties of the chains themselves was
discussed in numerous papers~\cite{i30,f1,f2,f3,f4,f5,f6,f7}.

The purpose of this work is to use the density
functional theory to calculate the force between two
plates covered with tethered charged brushes.
The approach used by us is an extension of the theory outlined
in reference~\cite{d3}. We will consider both the chains with the segments
having identical charges (i.e., typical polyelectrolytes) and
polyamphylic molecules with no net charge per chain.
In the latter case, we wonder how the distribution
of the charges along the chain (i.e., the chain architecture)
effects the solvation force and the structure of a system.

\section{Theory}

We consider two identical surfaces, lower and upper,  at a distance of $H$ apart.
Each  surface is covered with charged tethered chains
 $(P)$. The chains are built of $M$ tangentially jointed
charged hard spheres of the same diameter, $\sigma$.
The charge of the segment $j$, $j=1,\dots,M$ is  $Z^{(P)}_j e$, where $e$ is
the magnitude of an elementary charge.
The chain connectivity is ensured by imposing a bonding
potential~\cite{d0,d1,d2,d3}
\begin{equation}
 V_\mathrm{b}(\mathbf{R})=\sum_{j=i}^M v_\mathrm{b}(|\mathbf{r}_{j+1}-\mathbf{r}_j|),
\end{equation}
 where
$\mathbf{R}=(\mathbf{r}_1, \dots, \mathbf{r}_M)$ is the vector describing the positions of
all the segments. For tangentially bonded chains,
 the potential $V_\mathrm{b}$ satisfies the relation
\begin{equation}
 \exp\left[\frac{-V_\mathrm{b}(\mathbf{R})}{kT}\right]=
\prod_{j=1}^M \frac{\delta(|\mathbf{r}_{j+1}-\mathbf{r}_j|-\sigma)}{4\pi\sigma^2}\,.
\end{equation}

The first segment of each chain, $j=1$,  is pinned at one of the surfaces.
The surface  bonding potential has the form
\begin{equation}
\exp\left[\frac{-v^{(P)}_1(z)}{kT}\right] ={\cal C}\delta(z'),
\label{eq:1}
\end{equation}
where $\cal{C}$ is a constant and $z'=z-\sigma/2$ for lower and
$z'=z-(H-\sigma/2)$ for the upper surface.
 The interaction of all the remaining segments of a chain with the walls
is described by the potential
\begin{equation}
V_j^{(P)}(z)=v_{j}^{(P)}(z) + v_{j,\mathrm{el}}^{(P)}(z),
\label{eq:2}
\end{equation}
 where
$ v_{j}^{(P)}(z)$, $j=2,\dots,M$, is the hard-wall potential
\begin{equation}
 v_j^{(P)}(z)=\left\{
\begin{array}{ll}
 \infty, & \mathrm{for  \  \  } z < \sigma/2  {\rm \  \  or \ \ } z>H-\sigma/2, \\
0, & \mathrm{otherwise,  }
\end{array}
\right.
\label{eq:2a}
\end{equation}
and $v_{j,\mathrm{el}}^{(P)}(z) =V_{j,\mathrm{el}}(z)+V_{j,\mathrm{el}}(H-z)$, where
\begin{equation}
v_{j,\mathrm{el}}^{(P)}(z')=-\frac{4\pi}{\epsilon} Q Z^{(P)}_je^2 z'
\label{eq:2b}
\end{equation}
 is the Coulomb potential. In the above,
$Qe$ is the surface charge density of each wall and $\epsilon$ is the dielectric constant.
According to equation~(\ref{eq:2b}), the surface charge density on each wall is
assumed to be the same.
The total chain-surface potential is then $V^{(P)}(z_1,\dots,z_j)=
\sum_{j=1}^M v_{j}^{(P)}(z_j) +\sum_{j=1}^M v_{j,\mathrm{el}}^{(P)}(z_j)$.

The modified surfaces are in contact with a three-component fluid.  The  indices $\alpha=\mathrm{S}$,
$\mathrm{C}$ and $\mathrm{A}$  abbreviate solvent, cation and anion species,
respectively. The particles of the second and of the third species ($\mathrm{C}$,  $\mathrm{A}$) are
charged hard spheres as well, for the sake of simplicity, of a diameter $\sigma$,
and carry the charges $eZ_{1}^\mathrm{C}$ and $eZ_{1}^\mathrm{A}$.
In other words, we consider the so-called solvent primitive
model of an electrolyte. This is the simplest model of
electrolyte solutions that takes into account the
presence of non-zero size solvent molecules. The polar nature of a solvent is taken into
account by retaining the dielectric constant in the Coulomb
interaction between the ions. Note that the solvent
primitive  model was  used by Henderson et al.
to describe the forces between
macroscopic particles in an electrolyte and to model
electrolyte-membrane systems~\cite{s1,s2}.
Other examples of applications of this model can be
found in references~\cite{s3,s3_2,s4,s5,s6}

Despite its extreme simplicity, the solvent
primitive model has at least two virtues. It
may be as reasonable a model of a solvent as can be treated
conveniently by density functional theory and it does recognize that solvent molecules
occupy space.

In this work we
restrict our attention to the case of monovalent ions,
i.e., $Z_{1}^\mathrm{C}=|Z_{1}^\mathrm{A}|=1$ and $|Z_{j}^\mathrm{P}|=1$  for $j=1,\dots,M$.
 The solvent S molecules, however,
are  uncharged hard spheres of the same diameter, $\sigma$. Thus, $Z_1^{(\mathrm{S})}=0$.
The interactions of the fluid species with the surface are given by equations~(\ref{eq:2})--(\ref{eq:2b}),
in which $Z^{(P)}_j$ should be replaced by $Z^{(\eta)}_1$.

The interactions between all the spherical species (i.e., between the molecules S, A
and C and the consecutive segments of the chains P) are given by
\begin{equation}
u_{ij}^{(\eta\alpha)}(r)=\left\{
\begin{array}{ll}
\infty , & r<\sigma,\\
\frac{{e}^{2}Z_{i}^{\eta }Z_{j}^{\alpha }}{\varepsilon r}\,, & r>\sigma,%
\end{array}%
\right.  \label{eq:3}
\end{equation}
where $\eta,\alpha = \mathrm{S, C, A, P}$.

The confined system is in equilibrium with a three component bulk
mixture of the components  C,  A  and S.  The bulk system is at the same temperature
and at the same chemical potentials $\mu_{\eta}$, $\eta=\mathrm{A,C,S}$  as the confined system.
The bulk densities of particular species are $\rho^{\eta}_\mathrm{b}$.

In order to proceed, let us introduce the notation, $\rho^{(P)}(\mathbf{R})$ and $\rho^{(\eta)}(z)$,
$\eta=\mathrm{S, C, A}$
for the density distribution of chains and of fluid species, respectively.
The theory is constructed in terms of
the density of particular segments of chains, $\rho_{sj}^{(P)}(z)$,  and  the
total segment density of chains, $\rho_s^{(P)}(z)$. These densities are introduced via commonly used
relations~\cite{d0,d1,d2,d3,d5}
\begin{equation}
 \rho_{s}^{(P)}(\mathbf{r})=\sum_{j=1}^M \rho_{sj}^{(P)}(\mathbf{r})=\sum_{j=1}^M
 \int \rd\mathbf{R}\delta(\mathbf{r}_j-\mathbf{r})\rho^{(P)}(\mathbf{R}).
\label{eq:4}
\end{equation}
In the system under study, all the local densities are only  functions of the distance $z$ from the
lower surface that is set at $z=0$.

The system is studied in a grand canonical ensemble with the constraint of constancy of the
number of tethered chain molecules, i.e.,
\begin{equation}
\rho_\mathrm{c}=\int \rd z \rho_{sj}^{(P)}(z),
\label{eq:con}
\end{equation}
where $\rho_\mathrm{c}$ is the total number of tethered chain molecules per area of the surface.
Since the confined system comprises two surfaces, the
surface density of the chains is $R_\mathrm{C}=\rho_\mathrm{c}/2$.

The equilibrium density profiles are obtained by
minimizing the thermodynamic potential
\begin{equation}
 \mathcal{Y}= F+\int \rd\mathbf{R} \rho^{(P)}(\mathbf{R})\sum_{j=1}^Mv^{(P)}(z_j)+
\sum_{\eta=\mathrm{S,A,C}}\int \rd\mathbf{r}  \left[v^{(\eta)}(z)\rho^{(\eta)}(z) -\mu_{\eta}\right]
+\int \rd\mathbf{r}q(z)\Psi(z).
\end{equation}
In the above, $F$ is free energy functional and
$q(z)$ is charge density
\begin{equation}
\frac{ q(z)}{e}=\sum_{j=1}^M Z^{(P)}_j\rho^{(P)}_{sj}(z)+\sum_{\eta=\mathrm{S,A,C}}Z^{(\eta)}_1\rho^{(\eta)}(z).
\end{equation}

The electrostatic $\Psi(z)$ satisfies the Poisson equation
\begin{equation}
 \nabla^2\Psi(z)=-\frac{4\pi}{\epsilon}q(z).
\label{eq:poi}
\end{equation}
A solution of the differential equation~(\ref{eq:poi})
for a fluid confined between two walls is  given in reference~\cite{s2}. It requires denomination
of the value of electrostatic potential at the wall, $\Psi_0=\Psi(0)=\Psi(H)$.
From the electro-neutrality condition of the
system, it follows that
\begin{equation}
 Q+\int \rd z q(z) = 0.
\end{equation}

The principal task in applying the density functional theory is to derive an
expression for the Helmholtz energy as a functional of the
density profiles. Following previous works~\cite{iaa,ibb,icc,d1,d2,d3,d4},  the Helmholtz energy
is divided into an ideal term that depends on the
bond potentials and the architecture of polymers, and the excesses arising
from interactions of various forms. The latter
terms are responsible for the thermodynamic nonideality.
Within the framework of our model, the Helmholtz
energy functional, $F$,  is decomposed
into the following contributions,
\begin{equation}
 F= F_\mathrm{id}+F_\mathrm{ex}\, ,  \qquad F_\mathrm{ex}=F_\mathrm{P}+F_\mathrm{hs}+F_\mathrm{el}\,,
\label{eq:free}
\end{equation}
where  $F_\mathrm{id}$ is the ideal contribution,
\begin{equation}
 \frac{F_\mathrm{id}}{kT}= \frac{1}{kT} \int \!\!\rd\mathbf{R}\rho ^{(P)}(\mathbf{R})V_\mathrm{b}(\mathbf{R%
})+\int \!\!\rd\mathbf{R}\rho ^{(P)}(\mathbf{R})\left\{\ln \left[\rho ^{(P)}(\mathbf{R}%
)\right]-1\right\}.
\end{equation}
The volume exclusion (the hard-sphere)  term, $F_{\mathrm{hs}}$, is calculated
according to the Fundamental Measure Theory, cf.
references~\cite{iaa,ibb,icc,d0,d1,d2,d3},
\begin{eqnarray}
 \frac{F_\mathrm{hs}}{kT}&=&\int \rd\mathbf{r} \Bigg\{ -n_0\ln(1-n_3)+\frac{n_1n_2-
\mathbf{n}_{V1}\cdot \mathbf{n}_{V2}}
{1-n_3}\nonumber\\
&&{}+\frac{1}{36\pi}\Bigg[n_3\ln(1-n_3)+\frac{n_3^2}{(1-n_3)^2}\Bigg]
\frac{n_2^3-3n_2\mathbf{n}_{V2}\cdot\mathbf{n}_{V2}}{n^3_3}\Bigg\},
\end{eqnarray}
where $n_i$, $i=0,1,2,3$ and $\mathbf{n}_{Vj}$, $j=1,2$ are, respectively,
scalar and vector total weighted densities. The total weighted densities are
sums of the weighted densities of individual species. For example,
\begin{equation}
 n_i=n^{(P)}_i+n^\mathrm{(S)}_i+n^\mathrm{(A)}_i+n^\mathrm{(C)}_i,
\end{equation}
and
$n^{(P)}_i=\sum_{j=1}^Mn^{(P)}_{ij}$.
  Since the relevant equations defining the weighted densities have
been already presented in numerous
works~\cite{iaa,ibb,icc,d0,d1,d2,d3,d4}, we have omitted them here.

The contribution  $F_\mathrm{el}$, arising from the coupling between electrostatic and
hard-sphere interactions is written down employing the approach described
in detail in references~\cite{d3,rpm1} We have
\begin{equation}
 F_\mathrm{sl}=\int \Phi_\mathrm{el} (z) \rd z,
\end{equation}
where
\begin{equation}
  \Phi_\mathrm{el} (z) = -\frac{\sigma}{T^*}\left[ \sum_{j=1}^M \left(Z^{(P)}_j\right)^2\bar\rho^{(P)}_{sj}(z) +
 \sum_{\eta=\mathrm{A,C}}\left(Z^{(\eta)}_j\right)^2\bar\rho^{(\eta)}(z)\right] \frac{\Gamma}{1+\sigma\Gamma}+
\frac{\Gamma^3}{3\pi^2}\,.
\label{eq:freel}
\end{equation}
In the above $\Gamma=\left(\sqrt{1+2\kappa\sigma} -1\right)/2\sigma$, $T^*$ is the ``electrostatic''
reduced temperature, $T^*=kT\epsilon\sigma/e^2$,
\begin{equation}
 \kappa^2=\frac{4\pi}{T^*}\left[\sum_{j=1}^M \left(Z^{(P)}_j\right)^2\bar\rho^{(P)}_{sj}(z) +
\sum_{\eta=\mathrm{A,C}}\left(Z^{(\eta)}_j\right)^2\bar\rho^{(\eta)}(z)\right].
\end{equation}
The quantities $\bar\rho^{(P)}_{sj}$ and $ \bar\rho^{(\eta)}$
are the ``reference electrostatic system averaged densities'', which are calculated according to references~\cite{d3,rpm1,rpm2}.

Finally, the excess free energy functional due to the
intra-chain correlation is given by
\begin{equation}
 F_\mathrm{P}/kT={{\frac{1}{1-M}}}\left[\sum_{j=1}^{M-1}
 \int \rd\mathbf{r} n^{(P)}_0\xi^{(P)} \ln y_j  + \sum_{j=2}^{M}
 \int \rd\mathbf{r} n^{(P)}_0\xi^{(P)} \ln y_j \right] ,
\label{eq:fp}
\end{equation}
where $\xi^{(P)}=1-\textbf{n}_{V2}^{(P)}\cdot {V2}^{(P)}\big/\left(n_2^{(P)}\right)^2$ and
$y_j$ is the contact value of the cavity correlation function.
\begin{equation}
 y_j=\left\{ \frac{1}{1-n_3}+\frac{{{n_2\sigma\left[1-\mathbf{n}_{V2}\cdot
\mathbf{n}_{V2}/(n_2)^2\right]}}}
{4(1-n_3)^2}\right\}
\exp\left(-\frac{1}{T^*}\frac{Z_j^{(P)}Z_{j+1}^{(P)}(2\Gamma\sigma+\Gamma^2\sigma^2)}{(1+\Gamma\sigma)^2}\right).
\label{eq:cont}
\end{equation}

We are aware of the fact that in our treatment, the calculation of the values
of $\Gamma$ is performed without discriminating the ions belonging to chains and
free ions in the confined solution. In the light of the works~\cite{i2,i10,i22a,am1,am1a}, such ions should be distinguished. Therefore,
the above expressions (\ref{eq:freel})--(\ref{eq:cont}), and equation~(\ref{eq:cont}) in particular, are approximations.
Note that in the case of bulk fluids, the sums in equation~(\ref{eq:fp}) are identical.

At equilibrium  the density profiles minimize the thermodynamic potential $\mathcal{Y}$,
i.e.,
\begin{equation}
 \frac{\delta \mathcal{Y}}{\delta\rho^{(P)}(\mathbf{R})}= \frac{\delta \mathcal{Y}}{\delta\rho^{(\eta)}(\mathbf{r})}=0, \qquad \eta=\mathrm{S,A,C}\, .
\end{equation}
This condition leads to the equations
\begin{equation}
 \rho^{(P)}(\mathbf{R})=C\exp\left[ -V_\mathrm{b}(\mathbf{R})-\frac{1}{kT}\sum_{j=1}^M\lambda^{(P)}_j(z_j)\right]
\label{eq:den1}
\end{equation}
and
\begin{equation}
 \rho^{(\eta)}(\mathbf{r})=\exp\left[ -\mu_{\eta}/kT -\frac{1}{kT}\lambda^{(\eta)}(z) \right],
\qquad  \eta=\mathrm{S,A,C}\,,
\end{equation}
where
\begin{eqnarray}
 \lambda^{(P)}_j(z_j)&=&\frac{\delta F_\mathrm{ex}}{\delta \rho^{(P)}_{sj}(z_j)}+v_j^{(P)}(z_j)+eZ^{(P)}_j \Psi(z_j),\nonumber\\
 \lambda^{(\eta)}(z_j)&=&\frac{\delta F_\mathrm{ex}}{\delta \rho^{(\eta)}(z)}+v^{(\eta)}(z)+eZ^{(\eta)}\Psi(z), \qquad \eta=\mathrm{S,A,C}\,,
\end{eqnarray}
and where the constant C  is calculated from the normalization condition~(\ref{eq:con}).
The multidimensional density profile equation~(\ref{eq:den1}) can be then reduced to the equations
for local densities of consecutive segments, $\rho^{(P)}_{sj}(z_j)$ using the method
described in~\cite{d3}.

The solvation force per unit area is calculated from
\begin{equation}
 \frac{f_\mathrm{s}}{kT}=-\frac{\partial\mathcal{Y}(H)/AkT}{\partial H}-\frac{p}{kT}\,,
\end{equation}
where $A$ is the surface area and $p$ is the pressure of the bulk deference fluid involving
the components S, A and C.

\section{Results and discussion}

The systems in question are characterized by numerous parameters. In order to reduce their number,
all the calculations have been performed assuming constant total bulk density of the fluid, $\rho^*_\mathrm{b}=
\rho_\mathrm{b}\sigma^3=0.7$, where  $\rho_\mathrm{b}=\rho_\mathrm{b}^\mathrm{(A)} +\rho_\mathrm{b}^\mathrm{(C)} +\rho_\mathrm{b}^\mathrm{(S)}$.
 This value is close to the density of water solutions at standard temperature and pressure (STP) conditions.
The composition of the bulk fluid, i.e., the bulk mole fraction of the 1:1 electrolyte, $x=(\rho_\mathrm{b}^\mathrm{(A)}+\rho_\mathrm{b}^\mathrm{(C)})/\rho_\mathrm{b}$,
was varied.  Also, the value of the reduced electrostatic temperature was kept constant and equal to $T^*=0.2$.  This temperature
is lower than the temperature usually used in computer simulations of nonuniform electrolytes, but
our calculations have been intentionally carried out at a lower temperature in order to increase the role
of electrostatic interactions.
The parameters that characterize the brush are the number of segments and the surface
brush density, $R_\mathrm{C}^*=R_\mathrm{C}\sigma^3$.
 Numerous calculations were carried out  assuming that the total
charge of the brush is zero, i.e., that the number of positively and negatively charged
segments is the same, $M_+=M_-=M/2$. However, the distributions of ``$+$'' and ``$-$'' charges along
the chain were different. We have also studied the cases of all the segments of the brush
 bearing the same charges. Moreover, in some cases, we carried out calculations for
uncharged chains and for the chains with
non-zero resultant charge, lower than $Me$.
 In all cases, the electrostatic potential at the wall,
$\Psi_0$, was set to be zero.

We introduce the following codes to distinguish the systems under study. The symbol $m_1+n_1-m_2+n_2-,\dots$
abbreviates a chain whose first $m_1$ segments are positively charged, the next $n_1$~-- negatively charged, etc.
When the sequence of the charges along the chain is repeated, we use parentheses to group the repeating units,
for example the symbol $5(1+1-)$ means that the chain is built of 10 segments alternately charged with $+$ and $-$,
and whose first (pinned) segment is positively charged.

\begin{figure}[ht]
\begin{center}
\includegraphics[width=0.37\textwidth,clip]{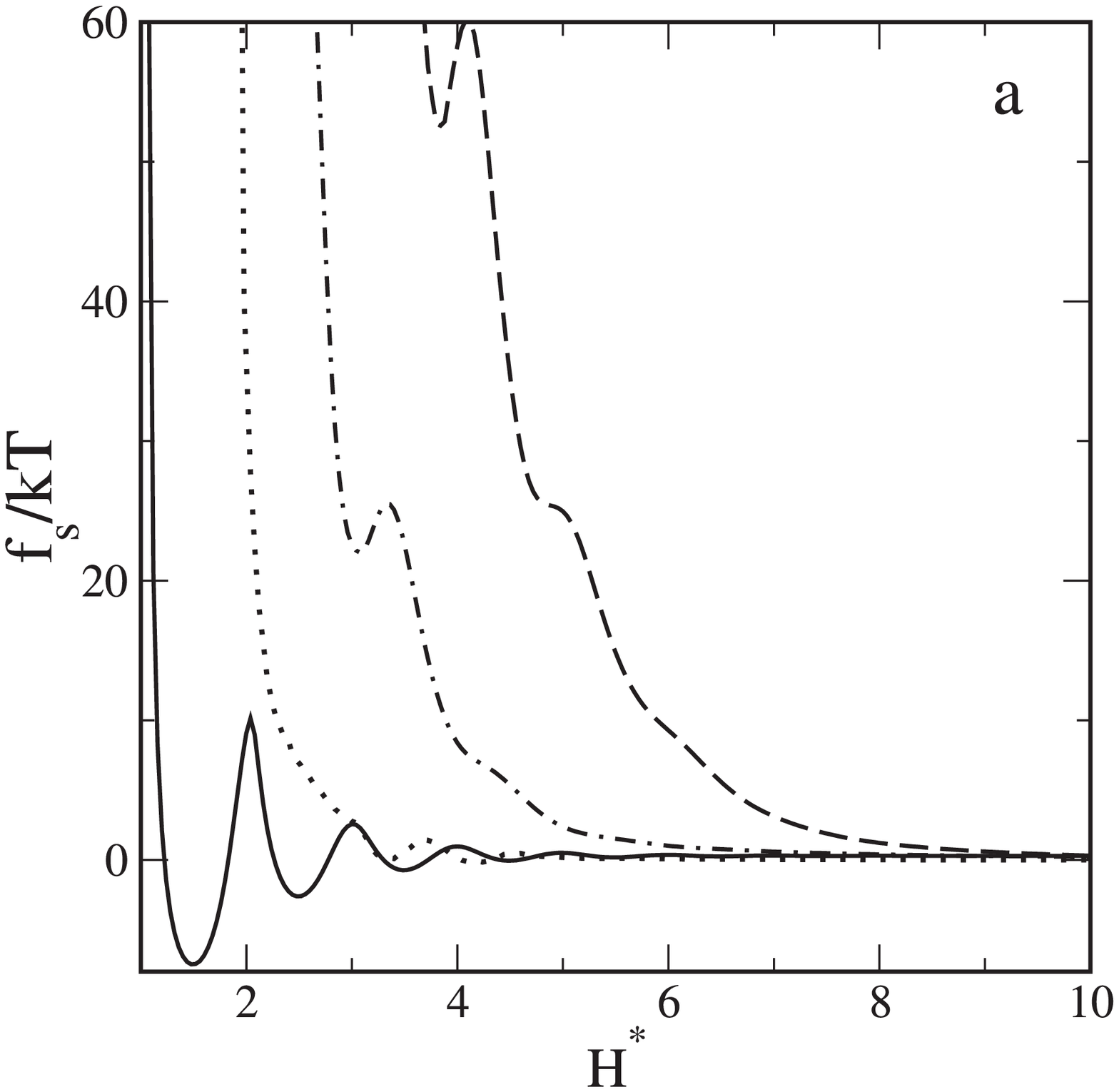}
\hspace{0.5cm}
\includegraphics[width=0.495\textwidth,clip]{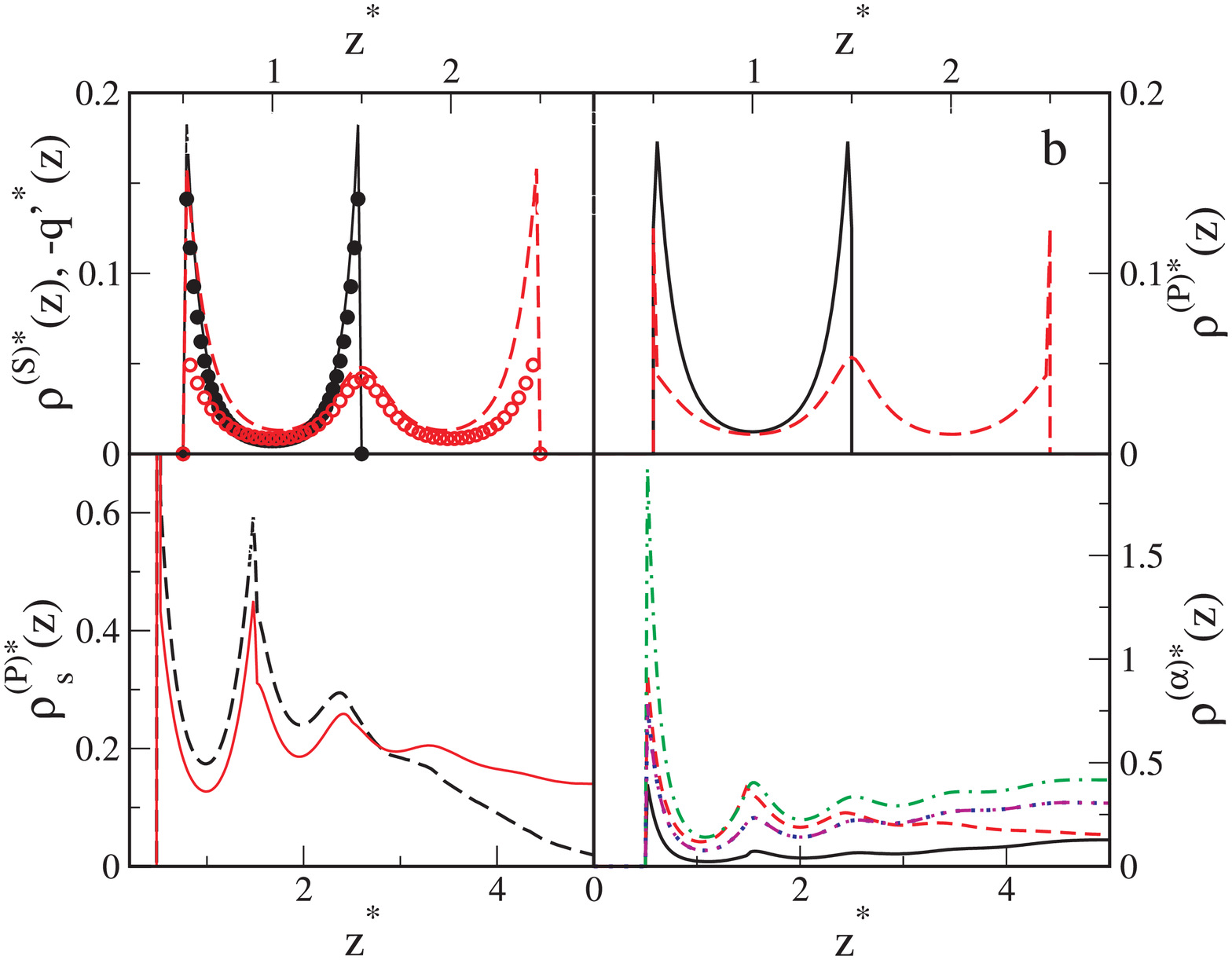}
\end{center}
\caption{(Color online) Part (a). The  solvation force vs.  $H^*=H/\sigma$.
All the results, except for those marked by a dotted line that were obtained for uncharged chains ($M=10$),
are for $10+$ chains.
Solid line is for $R_\mathrm{C}^*=0.0025$,  dash-dotted line~-- for $R_\mathrm{C}^*=0.1$ and
dashed line -- for $R_\mathrm{C}^*=0.15$.
Part~(b). Examples of the local densities of the chains $10+$ and of the fluid.
Upper panels are for $R_\mathrm{C}^*=0.0025$ and $H^*=3$ and 2, while lower panels~-- for
$R_\mathrm{C}^*=0.1$ and $H^*=10$. Left upper panel: solvent density profiles
and the profiles of charges, due to ionic species. Right upper panel:
the total segment density profiles. The calculations are for $H^*=3$ (dashed lines and open symbols)
and for $H^*=2$ (solid lines and filled symbols).
Lower left panel: the total segment density profiles of charged (solid line) and of
uncharged chains (dashed line). Right lower panel: the profiles of positive (dashed line),
negative (dash-dotted line) ions and solvent (solid line, the latter profile is multiplied by 10) for
 $10+$ brush. For uncharged brush, the  ionic species of profiles are identical
(dash-doubly dotted line). The solvent profile is given by a dotted line. The latter profiles
are multiplied by 10. The bulk mole fraction of the electrolyte is
$x=0.1$.}
 \protect
\label{fig:1}
\end{figure}
In figure~\ref{fig:1}~(a), we show how the solvation force depends on the surface density of the grafted chains  $10+$, whose  all segments bear the unit positive charges. For a comparison, we also display here the result for uncharged
chains built of $M=10$ segments.
For a low surface density of the chains, $R_{\mathrm{C}}^*=0.0025$,  the
course  of the solvation force vs. $H^*$ reminds
us the course for the pore with non-modified walls~\cite{iaa,icc}.
The solvation force diverges at low wall-to-wall separations $H^*=H/\sigma$ and exhibits oscillations corresponding to attractive and
repulsive forces that are well-pronounced at small values of $H^*$ and decay to zero at larger plate-to-plate
separations.  The maxima of the solvation force correspond
to the development of consecutive layers of the solvent density profiles, see figure~\ref{fig:1}~(b). The role of
chains and of electrostatic interactions in particular is small and the solvation force is
almost entirely determined by the solvent packing effects.

When the surface density of chains increases, the dependence of the  solvation force on $H^*$ quantitatively
 changes. For charged chains ($10+$), the solvation force becomes repulsive for all values of $H^*$.
Two factors play a significant role here. The first one is the volume exclusion effect: with an increasing  $R_\mathrm{C}$, the segments of the chains occupy more and more space, and further compression
becomes difficult for small  $H^*$. The second factor is electrostatic repulsion between the segments of chains pinned
to the opposite walls.  For $R_\mathrm{C}^*=0.1$, the second effect becomes very important. Indeed,
the solvation force for charged and uncharged brush of the surface density $0.1$ is very different.
For uncharged chains [dotted line in figure~\ref{fig:1}~(a)], the solvation force becomes even slightly attractive (e.g., for $H^* \approx 4$).
For $H^*\leqslant 3$ the solvation force becomes repulsive and grows rapidly.  Electrostatic forces between the segments cause the appearance of
strong repulsion between two walls at larger distances $H^*$.
The above results are in quantitative agreement with experiments~\cite{e2,f3}.

Figure~\ref{fig:1}~(b) shows examples of a structure within some selected pores. Upper panels are for
$R_\mathrm{C}^*=0.0025$ and for the chains $10+$.  Left upper panel displays the solvent density profiles (solid lines)
and the negative total charge profiles (symbols) for two pores with $H^*=3$ and $H^*=4$.
For $H^*=4$, three well-developed layers of the solvent appear within the pore. This structure is
resistant to compression and a local maximum
appears on the plot of the solvation force. When $H^*$ decreases down to 3, the inner layer is ``squeezed out'' the pore, and
the next structure develops which is resistant to further compression.  The total segment density profiles
[the right upper panel of figure~\ref{fig:1}~(b)] behave similarly to the profiles of the solvent.

The average density of a fluid within the pore is much lower than in the bulk
reference system (we recall that the latter equals 0.7). A part of the space inside the pore
is occupied by the segments of chains [see the upper right panel of figure~\ref{fig:1}~(b)], but
electrostatic interactions cause a lower average density in the confined system that is in chemical equilibrium
with the bulk system. The
upper left panel shows the negative charge density profile of
ionic species of the fluid, $-{q'}^*(z)=-q'(z)\sigma^3/e$, where
$q'(z)=\sum_{\alpha=\mathrm{A,C}}\rho^{(\alpha)}(z)$.

Despite the fact that the electrostatic potential at the pore walls is zero, $\Psi_0=0$,
an effective negative charge on the pore walls is generated. This is the result of the charges on the  segments.
Unlike the pores with non-modified walls ($R_\mathrm{C}^*=0$), the zero surface charge does not correspond
to the zero of the electrostatic potential at the wall. It would be of interest to determine the
dependence of the so-called ``potential of zero charge'', PZC, on the
pinned chain characteristics,
but this problem is out of the scope of the current research.

\begin{figure}[!b]
\begin{center}
\includegraphics[width=0.5\textwidth]{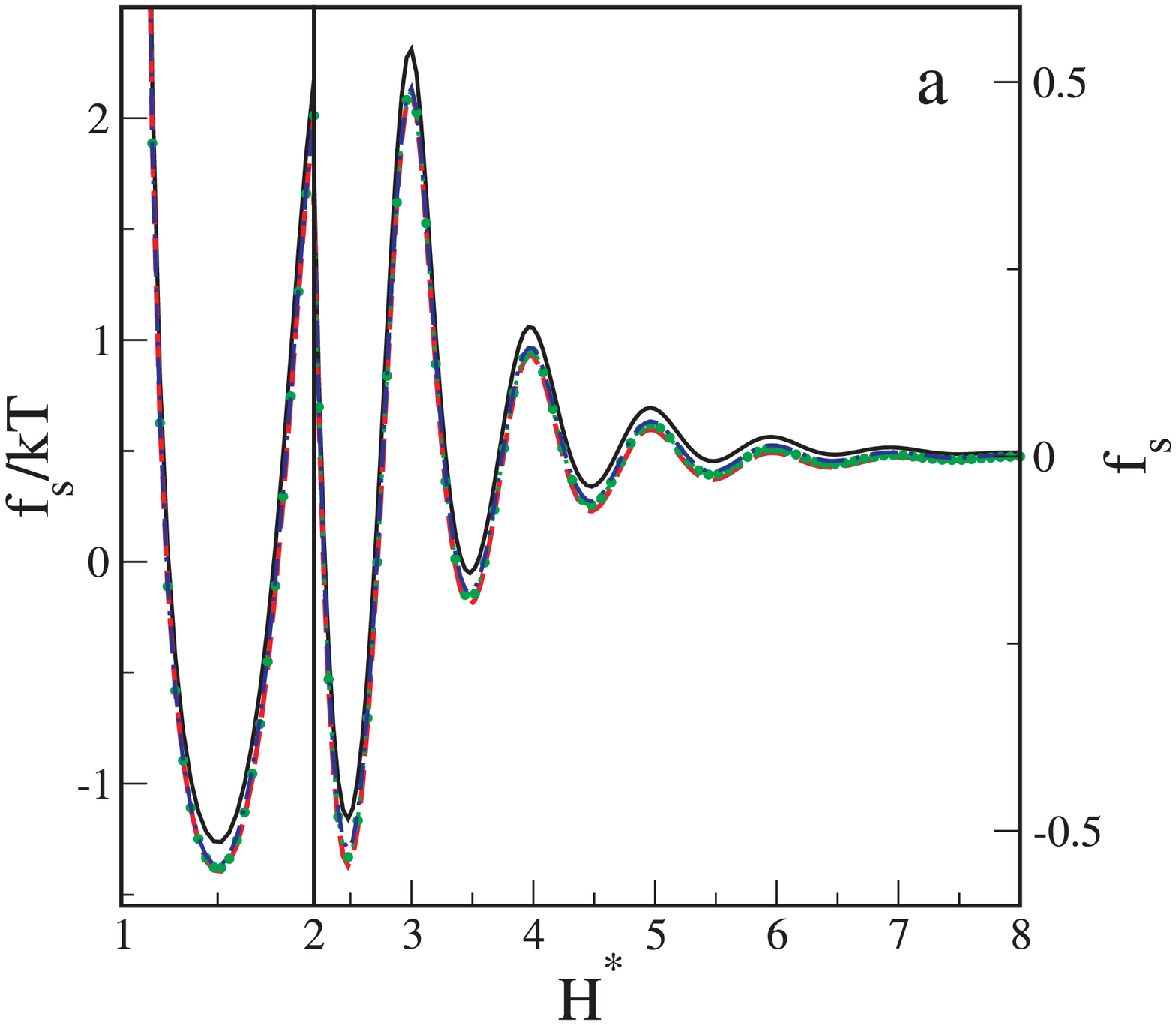}
\hfill
\includegraphics[width=0.44\textwidth]{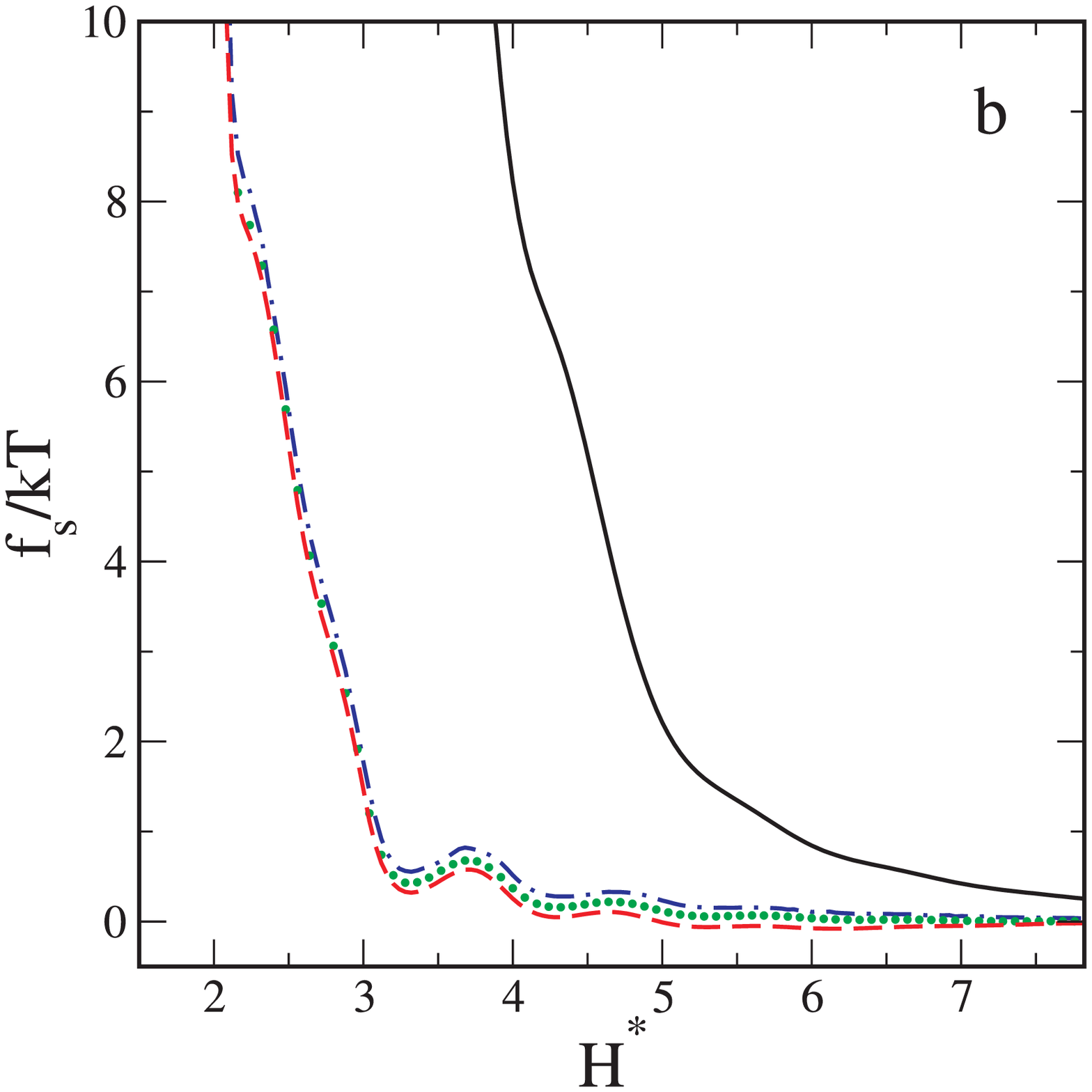}
\end{center}
\caption{(Color online) The solvation force vs. $H^*=H/\sigma$.
Solid line is for the $10+$ chains, dashed line -- for the chains $5(1+1-)$,  dotted line -- for
$4(2+2-)1+1-$ and dash-dotted line -- for $2(5+5-)$. Part a is for $R_\mathrm{C}^*=0.01$ and part b --
for $R_\mathrm{C}^*=0.1$.
In all cases the bulk mole fraction of the electrolyte is
$x=0.1$.}
 \protect
\label{fig:2}
\end{figure}

Lower panels in figure~\ref{fig:1}~(b) show the structure inside the pore of $H^*=10$ wide. The grafting density
if high, $R_\mathrm{C}^*=0.1$. Due to the symmetry, only
one half of the profiles is shown. We compare here the situation for uncharged
chains built of $M=10$
segments and for the chains $10+$. Uncharged tethered
chains are much more coiled. Indeed, their density
is low at the pore center. The stretching of charged chains is caused by electrostatic repulsion between the segments.
Positively charged segments attract negative ions, which, in turn, cause the accumulation of
positive ions. However, the concentration of solvent molecules inside a pore is very low, because there is simply ``no room'' for them. In the case of uncharged chains, the concentration of all components of the fluid is very low inside the pore. This is because there is no  attraction between the segments and  ions,
and the volume exclusion effects are responsible for a very
low confinement of all the fluid species.

Here we consider  the cases of the total charge of a single chain
 equal to 0. We studied the following  distributions of
the charges along the chain:
(a) $5(1+1-)$, (b) $4(2+2-)1+1-$ and (c) $2(5+5-)$. The brush $10+$ is treated as a reference.
Figure~\ref{fig:2}~(a) shows the solvation forces for  four systems  defined above. The surface
grafting density was $R_\mathrm{C}^*=0.01$.
Surprisingly, despite  quite big differences between the considered
chains (chains with the total net charge, but with different distribution of charges vs. the chain with the total charge
equal to  $10e$), the differences between the solvation force for particular systems are rather small.
The solvation force for the chain $10+$ is more repulsive. The maxima of $f_\mathrm{s}$ are higher and the minima
are shallower than for all the remaining cases, as expected from
the consideration of the role of electrostatic forces.
Quite unexpected, however, is a very small effect of  different distributions of the charges along the chain
on the solvation force. Indeed,
the results for the systems (a)--(c) are hardly distinguishable
on the figure scale.
At a higher grafting density, $R_\mathrm{C}^*=0.1$,  the solvation
forces for the systems (a)--(c) are still very similar. However, the  solvation force for the chains $10+$
is very different from all the remaining curves and shows that an electrostatic repulsion occurs
at larger wall-to-wall separations,  cf. figure~\ref{fig:2}~(b).
The pinned chains (a)--(c) can assume coiled configurations at
both pore walls, whereas for the chain $10+$, electrostatic repulsion between
segments inhibits its coiling.

\begin{figure}[!h]
\begin{center}
\includegraphics[width=0.48\textwidth,clip]{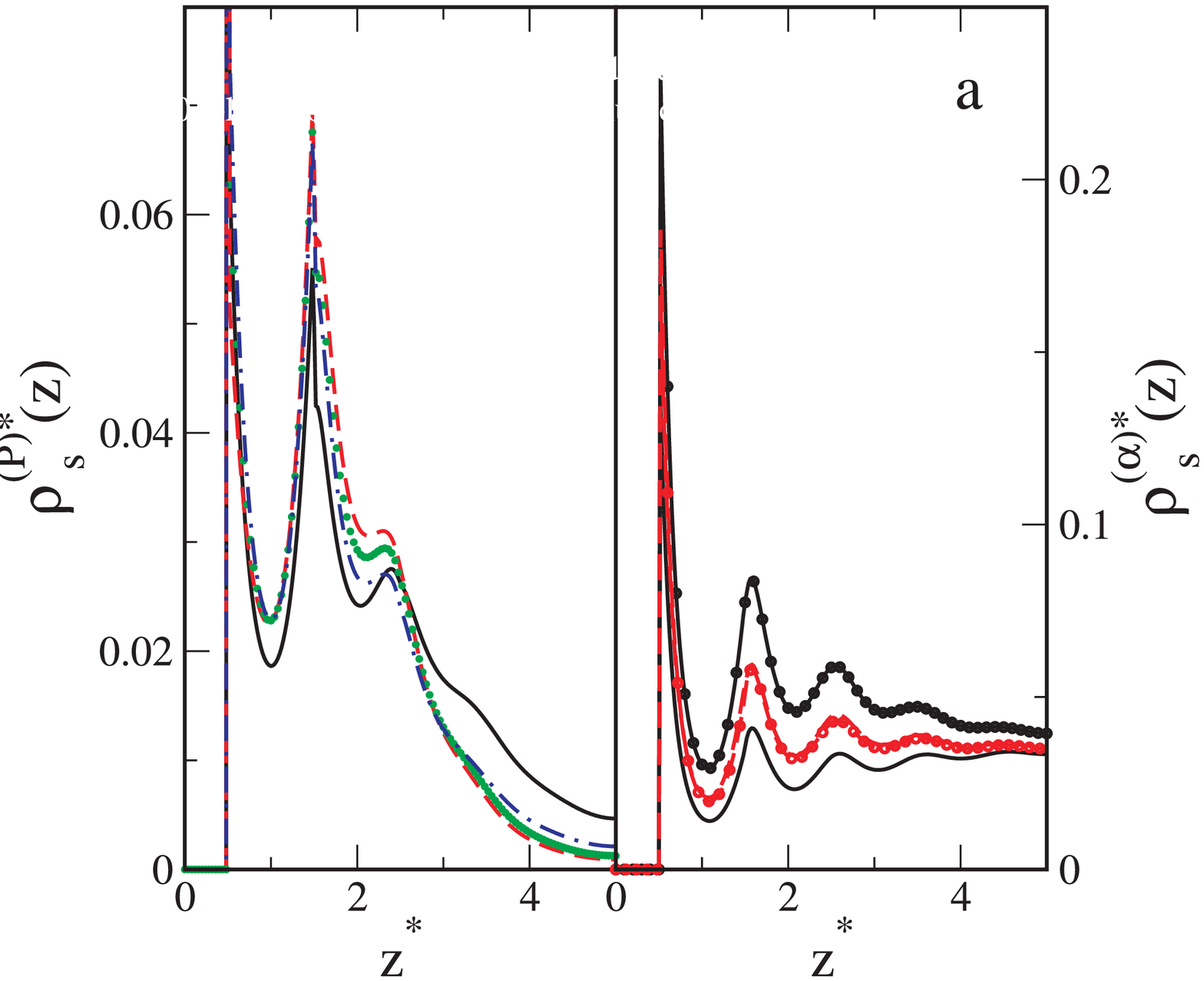}
\hfill
\includegraphics[width=0.48\textwidth,clip]{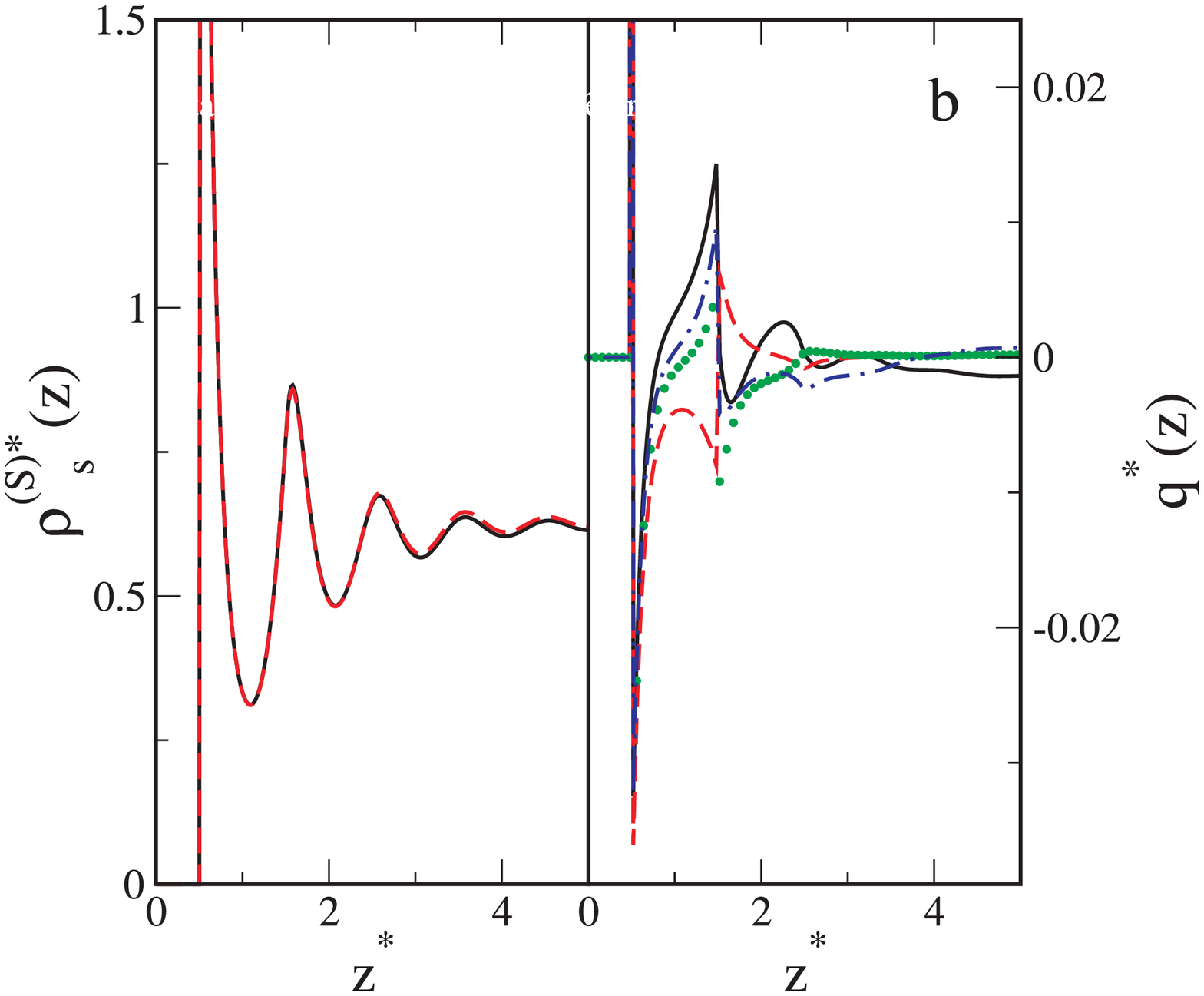}
\end{center}
\caption{(Color online) Part (a). Left panel shows the total segment  density profiles
for the systems in figure~\ref{fig:2}~(a). Abbreviations are the same as in figure~\ref{fig:2}~(a).
Right hand panel shows the profiles of ions for two systems shown in figure~\ref{fig:2}~(a), namely
for $10+$ (solid lines) and $5(1+1-)$ (dashed lines). Lines without symbols are for anions,
$\alpha=\mathrm{A}$,
lines with symbols are for cations, $\alpha=\mathrm{C}$.
Part (b). Left hand panel shows the solvent density profiles for grafted chains  $10+$ (solid line) and
$5(1+1-)$ (dashed lines), while the right hand panel shows the total charge density profiles, $q^*(z)=q(z)\sigma^3e$.
Abbreviations are the same as in figure~\ref{fig:2}~(a).  Due to the symmetry, only one half of the profiles
is displayed. The pore width is $H^*=10$, and the bulk mole fraction of an electrolyte is $x=0.1$.
}
 \protect
\label{fig:3}
\end{figure}

Examples of a structure  for the systems shown in figure~\ref{fig:2}~(a) are shown in figure~\ref{fig:3}.
Although the width of a pore is quite large, $H^*=10$, there is
still a significant overlap of the chains
pinned at the opposite walls, especially for the $10+$ chains.
Despite  small differences between the solvation forces
[figure~\ref{fig:2}~(a)], the differences between the total segment density profiles [right panel in figure~\ref{fig:3}~(a)]
for the cases (a)--(c) and $10+$ are pronounced. Although the latter chains are more stretched  than all
the remaining chains, the total segment density profile for the chains $10+$ is
rather low at the pore center. Due to a low surface density, there is enough space for
the segments to assume a slightly tilted or parallel to the wall configuration. Right
panel in figure~\ref{fig:3}~(a) shows the profiles of ions, $\alpha=\mathrm{A,C}$. A significant difference
between the profiles of anions and cations is observed for the chains $10+$ due to
electrostatic segment-anion attraction. For
electro-neutral  chains, the differences between the profiles  of anions and cations are
small.  However, the profiles of ionic species are not identical. This indicates that
there should be a small extra charge on the pore walls. In other words, the PZC for electro-neutral chains
is not at $\Psi_0=0$.  An interesting question is how PZC depends on the architecture of the
chains, but this problem is out of scope of the current research.

Figure~\ref{fig:3}~(b) shows examples of the solvent density profiles (right hand panel) and
the charge density profiles (left hand panel). Distribution of the solvent
particles inside the pore depends very little on the charges of the segments.
The solvent particles ``do not feel'' the charges, and the volume exclusion effects
almost completely determine the structure of the confined solvent.
The total charge distribution is sensitive to the architecture of the
grafted chains.

\begin{figure}[ht]
\centerline{
\includegraphics[height=0.4\textwidth,clip]{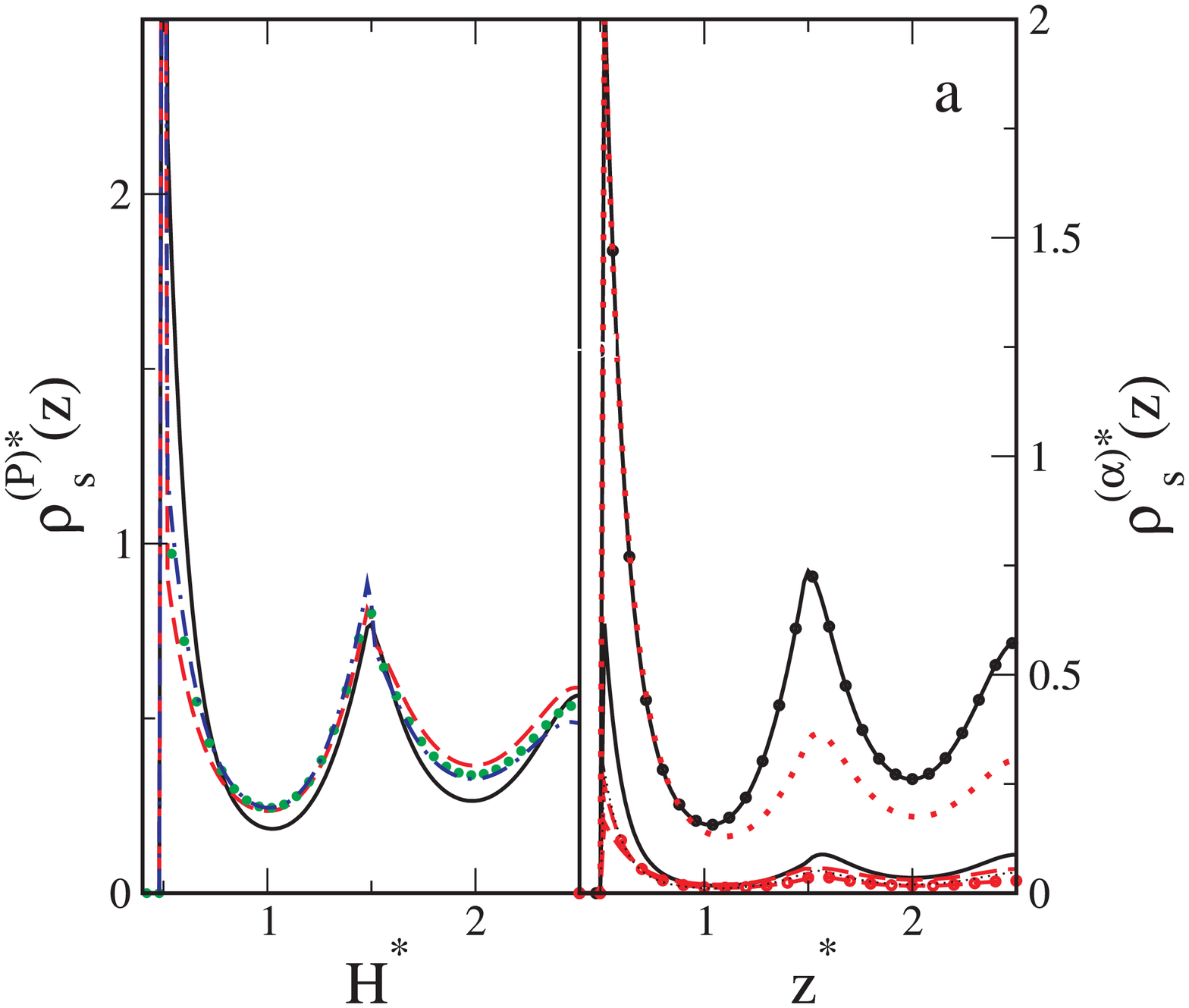}
\hfill
\includegraphics[height=0.4\textwidth,clip]{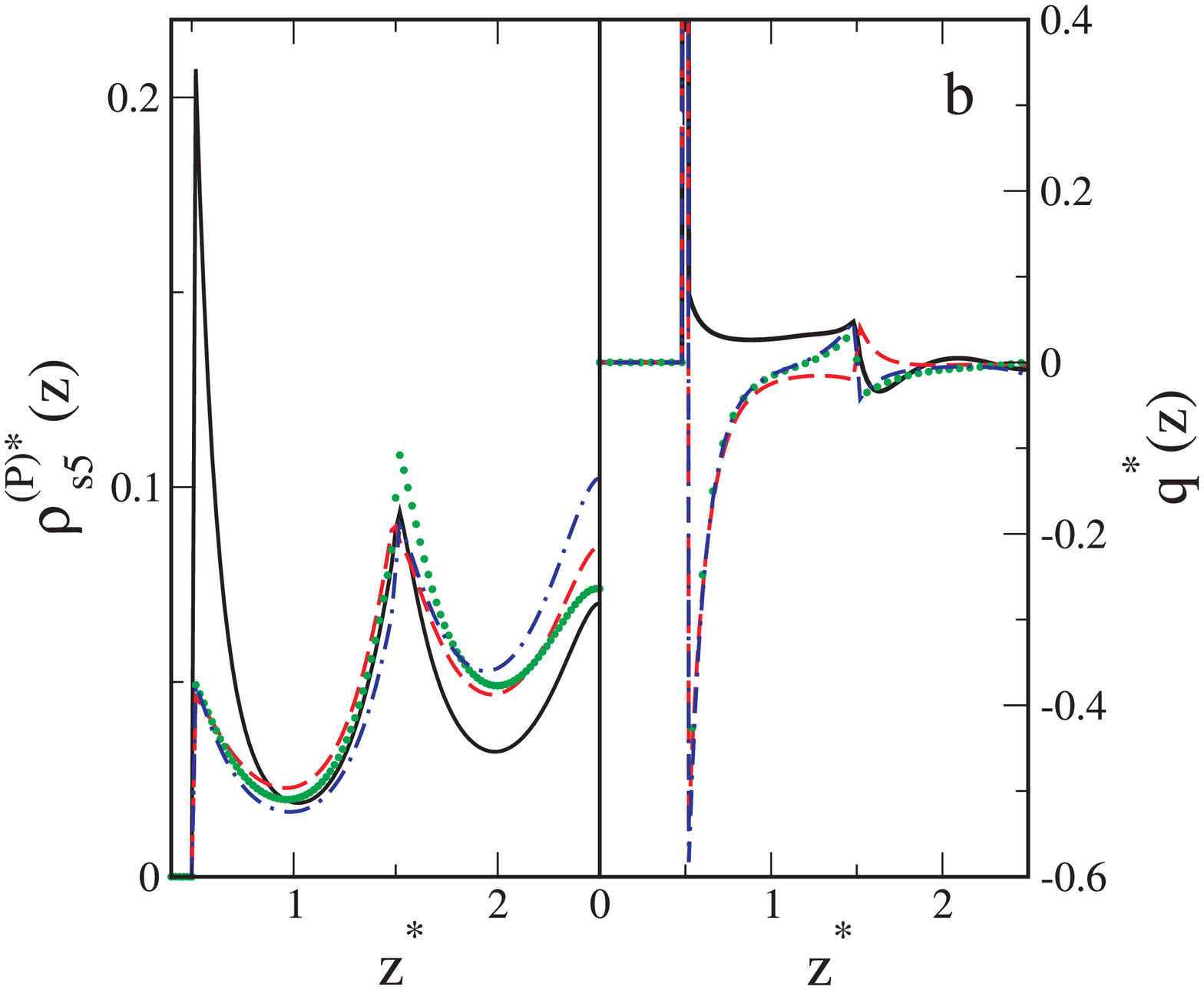}
}
\caption{(Color online) Part (a). Left panel shows the total segment  density profiles
for the systems in figure~\ref{fig:2}~(b). Abbreviations are the same as in figure~\ref{fig:2}~(b).
Right panel shows the profiles of ions for two systems from figure~\ref{fig:2}~(b), namely
for $10+$ (solid lines) and $5(1+1-)$ (dashed lines). Lines without symbols are for anions
$\alpha=\mathrm{A}$, with symbols~--
for cations, $\alpha=\mathrm{C}$. The latter profile was multiplied by 100 to make it visible. We also shown here the solvent profile
for the system   $5(1+1-)$ (dotted line, $\alpha=S$).
Part (b).   The profiles $\rho^{(P)}_{s5}(z)$ (left panel)
and the total charge profiles (right panel) for the systems from figure~\ref{fig:2}~(b). The meaning
of the lines is the same as in figure~\ref{fig:2}~(b).
The pore width is $H^*=5$ and the bulk mole fraction of the electrolyte is $x=0.1$.
}
 \protect
\label{fig:4}
\end{figure}

Similar investigations of the structure have also been carried out
for a higher grafting density, $R_\mathrm{C}^*=0.1$. The solvation forces for those systems
are shown in figure~\ref{fig:3}~(b). Figure~\ref{fig:4}~(a) compares the total segment density profiles
for the systems $10+$ and for the systems (a)--(c) (left hand panel), as well as the density profiles of fluid components
for the systems $10+$ and $5(1+1-)$ (right hand panel). The  pore is narrow, $H^*=5$, so the segments
of the chains are  compressed and form  layered structures. The differences between
the total segment densities for different systems are not big.  Because of a small pore width,
there is not much room for different rearrangements of the segments
and the observed layered structure permits to effectively
pack the segments inside the pore.

We have also inspected the density profiles of individual segments, but for the sake of
brevity only the profiles of the middle segments, $i=5$, are shown in the
left hand panel of figure~\ref{fig:4}~(b). The differences between these profiles are significant.
The performed analysis of the shape of the functions
$\rho^{(P)}_{si}(z)$, $i=1,2,\ldots,10$ leads to the conclusion
that in the case of  $10+$, the
layers of chains pinned
 at the opposite pore walls  penetrate into one another.
However, in
the remaining  (a)--(c) cases, the chains
 ``turn back'' at the middle of the pore and the
inter-penetration of the layers attached to the opposite walls is
much weaker.

In the case of the pore $H^*=5$ modified with  $10+$ chains,
 the confined fluid
contains almost solely anions  (figure~\ref{fig:4}~(a), left hand panel). The concentration of cations is extremely low;
in order to make the cations profile visible on the figure scale, we multiplied it by 100. Similarly,
the density of the solvent inside the pore is also extremely low. In the case of a pore
modified with $5(1+1-)$ brush, the anions accumulate at the pore walls, while
in the pore interior cations prevail.
Moreover, for $5(1+1-)$ chains, we also observe
a concentration of solvent molecules inside the pore.

The shape of total charge distributions inside the pore
exhibits quite large local changes.
 In all the cases studied, there appears
a very high peak at $z^*=0.5$ on the curves $q(z)$ (we have cut the height of this
peak on left hand panel of figure~\ref{fig:4}~(b) in order  to make the figure readable). The height of this peak  almost
exactly corresponds to the amount of pinned charged segments (in all the cases, the
pinned segments are positively charged).  For a system $10+$, the function $q(z)$ is positive
within the region of  $0.5<z<1.5$, while for the same region it is negative for all the remaining systems in
question.  Similarly to the
previously considered cases, the integrals of $q(z)$  over the entire pore are different from zero, i.e.,
there should be extra charges at the pore walls in order to ensure the total electro-neutrality of the entire system.
In other words, the PZC does not correspond to the zero value of  electrostatic potential at the wall.

\begin{figure}[!h]
\begin{center}
\includegraphics[height=0.45\textwidth,clip]{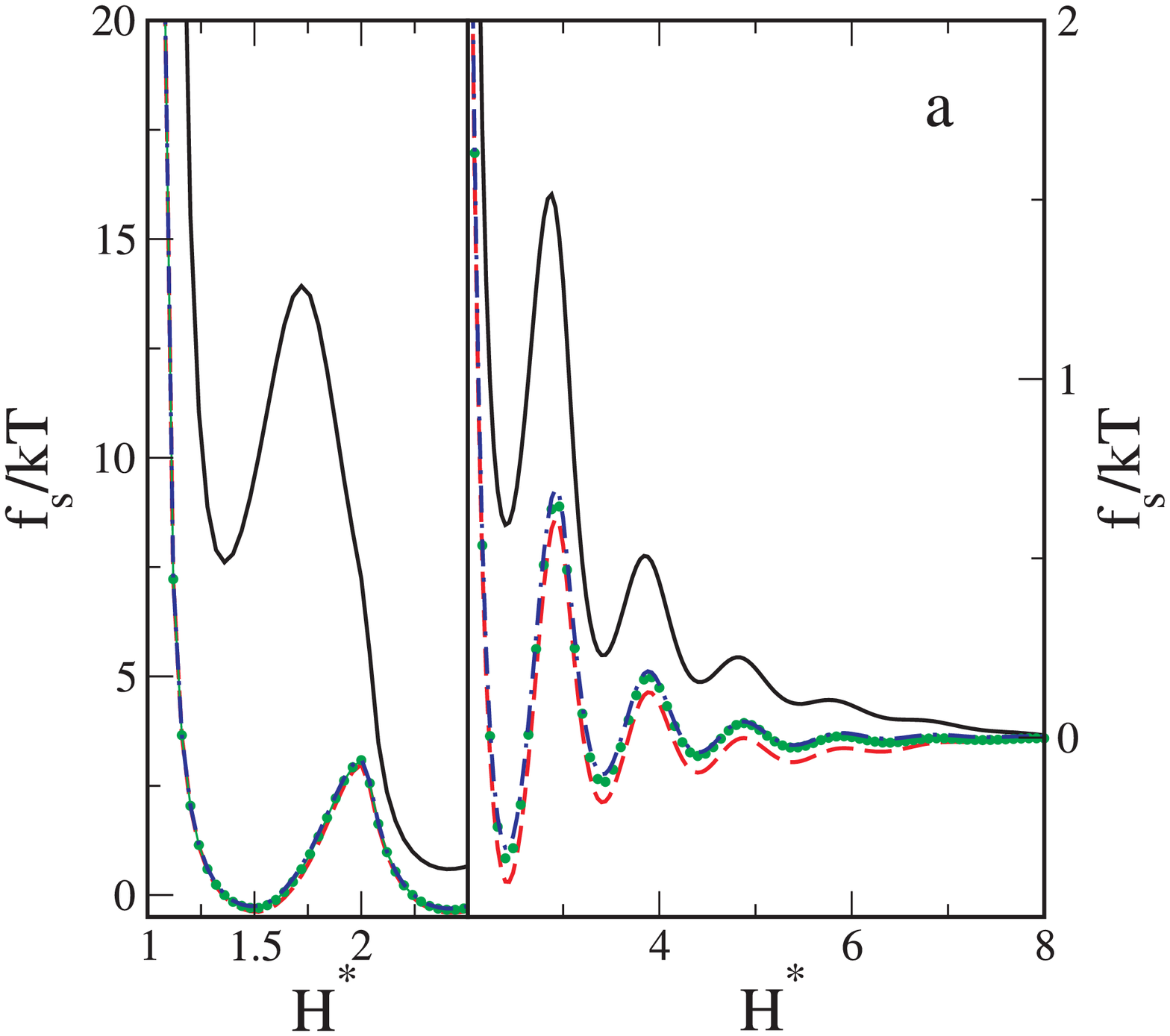}
\hfill
\includegraphics[height=0.45\textwidth,clip]{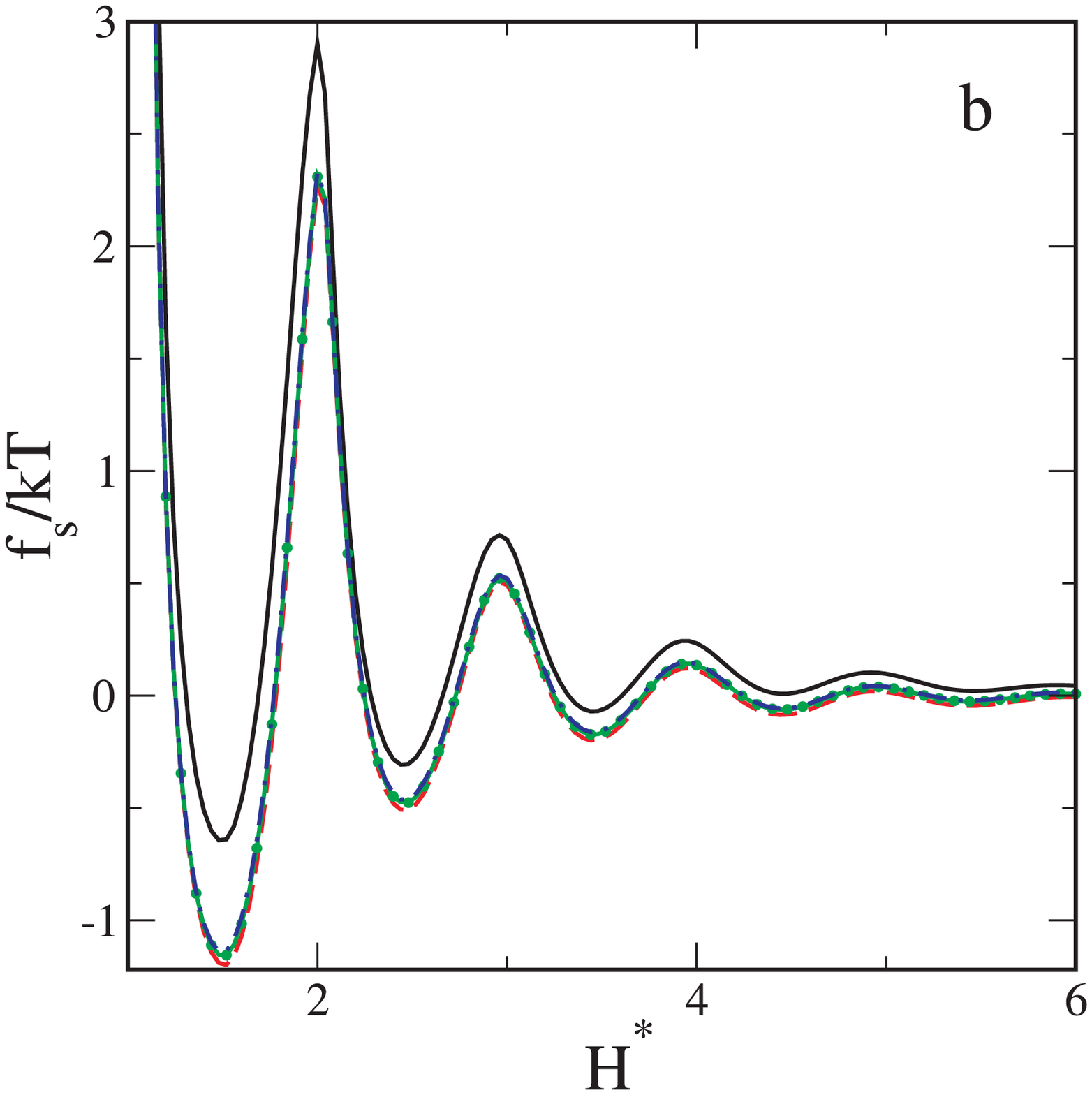}
\end{center}
\caption{(Color online) Solvation force for the brushes $8+$ (solid line),  $4(1+1-)$~--dashed line,
$2(2+2-)$~-- dots and  $4+4-$ and dash-dotted line. Part~(a) is for $R_\mathrm{C}^*=0.05$, part~(b) is
for $R_\mathrm{C}^*=0.025$. The bulk mole fraction of the electrolyte is $x=0.1$.
}
 \protect
\label{fig:5}
\end{figure}
\begin{figure}[!h]
\begin{center}
\includegraphics[height=0.45\textwidth,clip]{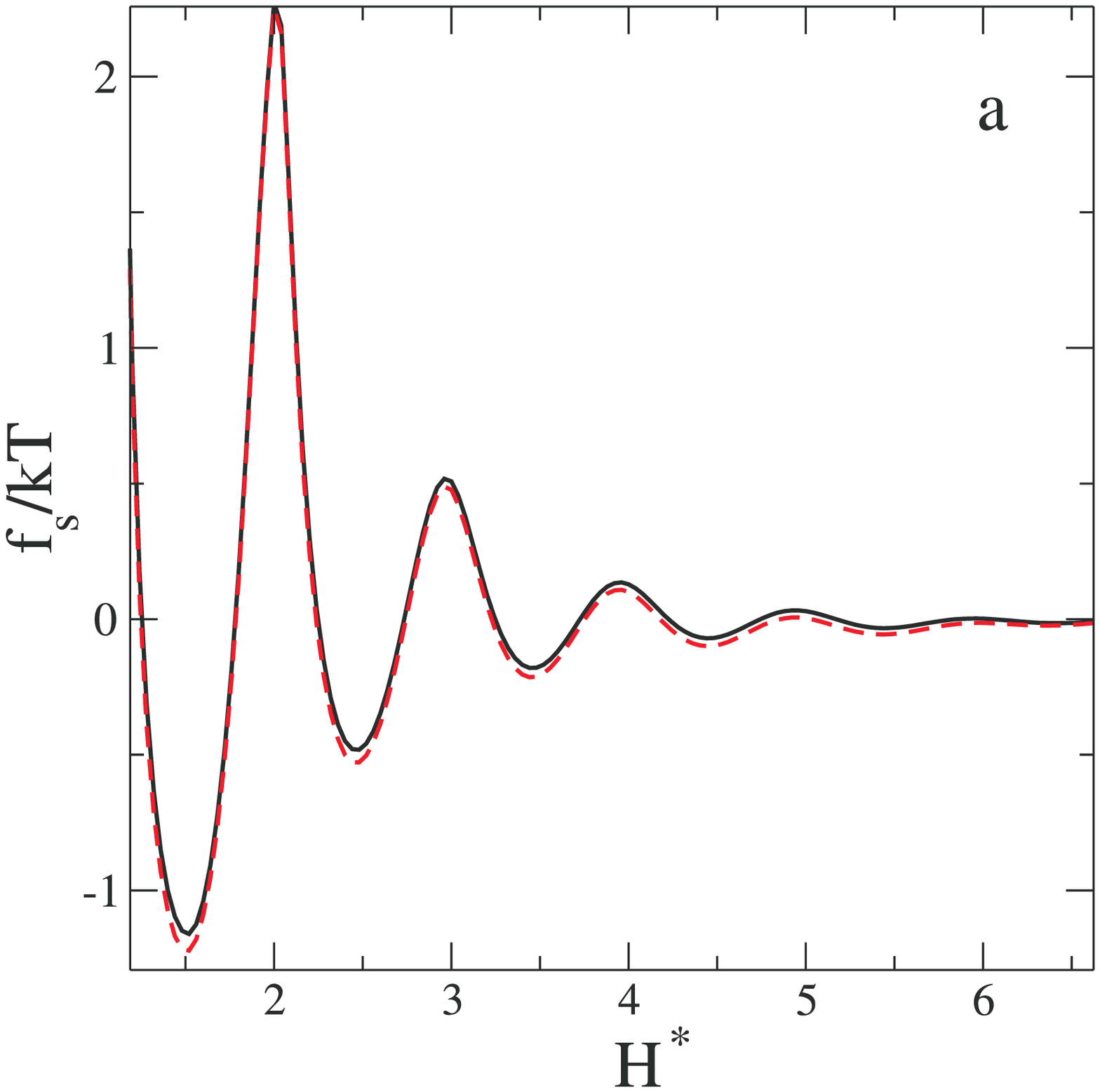}
\hspace{1.5cm}
\includegraphics[height=0.42\textwidth,clip]{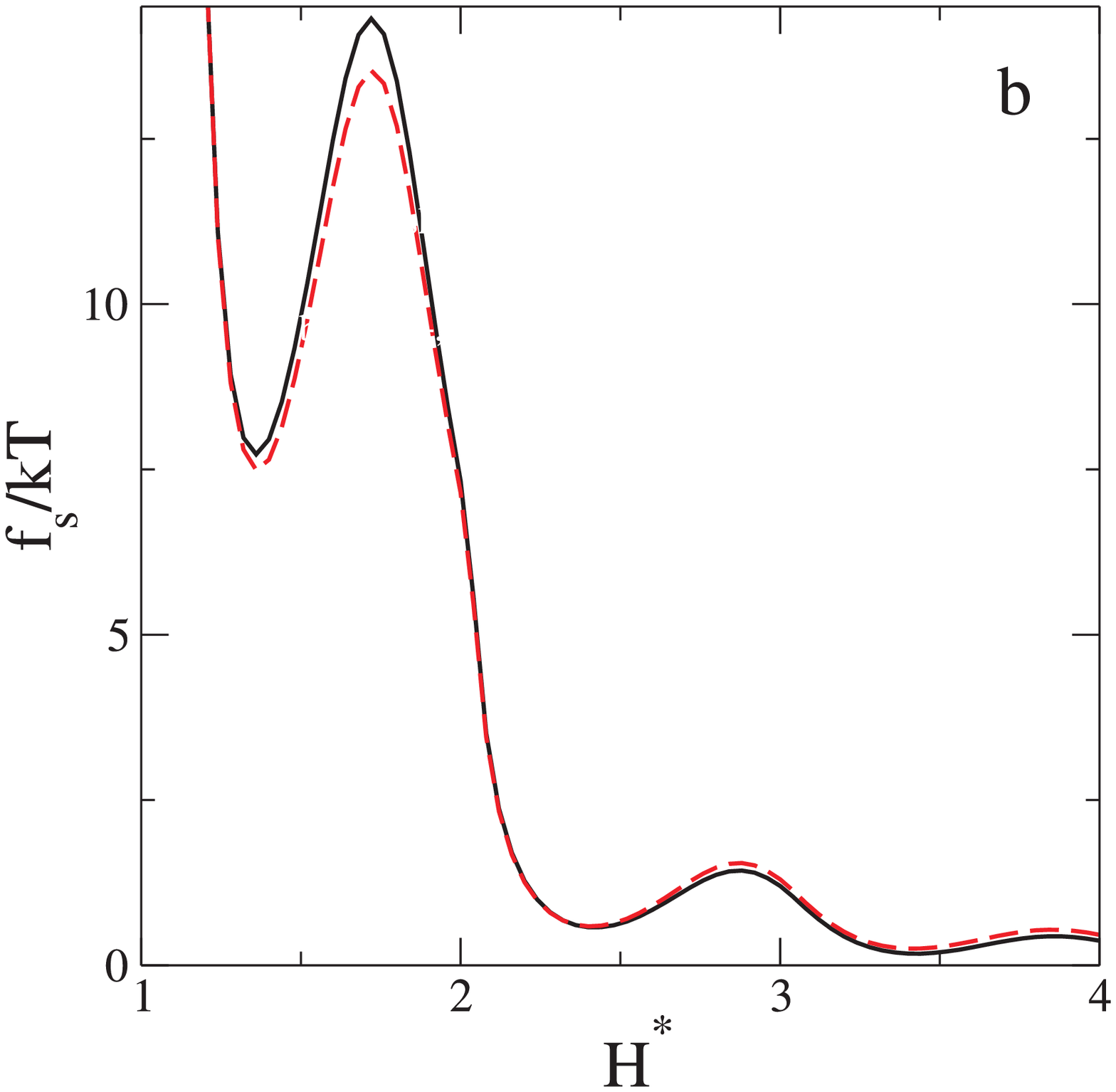}
\end{center}
\caption{(Color online) The dependence of the solvation force for the systems
 $4(1+1-)$ [part~(a)]
and $8+$ [part~(b)] on the bulk mole fraction of the electrolyte. Solid line is for $x=0.2$, while dashed line is for  $x=0.05$.  The grafting density is $R_\mathrm{C}^*=0.025$ in part~(a) and 0.5 in part~(b).
}
 \protect
\label{fig:6}
\end{figure}

Now we consider the chains built of $M=8$ segments. Similarly,  we study the
case when all the segments are positively charged, $8+$, as well as three electro-neutral chains: (a) $4(1+1-)$;
(b) $2(2+2-)$ and (c) $4+4-$.  The calculations of the solvation force  have been
carried out for several grafting densities $R_\mathrm{C}^*$, but only some selected results
are shown in figure~\ref{fig:5}. In general, all the features observed for the chains
 of 10 segments
 also appear here.  When the grafting density is not too high,
the solvation force oscillates around zero. For higher values of $R_\mathrm{C}^*$, the solvation force
becomes entirely repulsive, especially in the case of chains $8+$. For electro-neutral chains,
the solvation force only weakly depends on distribution of the charges
along the chain.  In the latter case, the difference between the solvation force for
different architectures of the brush slightly increases with an increase of
the grafting density [cf. figure~\ref{fig:5}~(a) and figure~\ref{fig:5}~(b)].  Since the changes in the structure of a confined
fluid with distribution of charges along the chain are
quantitatively similar to $M=10$ cases, the relevant figures have been omitted.

Figure~\ref{fig:6} shows the dependence of the solvation force on
 the bulk mole fraction of an electrolyte. The calculations are for two
types of grafted chains, namely for $4(1+1-)$ [part~(a)]
and for $8+$ [part~(b)]. In both cases, the effect of the bulk concentration of ions
on the solvation force is small. It is slightly more pronounced for $8+$
chains, especially
for t $1.5<H^*<2$. In other words, the most important factors effecting the solvation
force are the characteristics of the chain, i.e., its length, grafting density
and architecture.

\begin{figure}[!ht]
\begin{center}
\includegraphics[height=0.45\textwidth,clip]{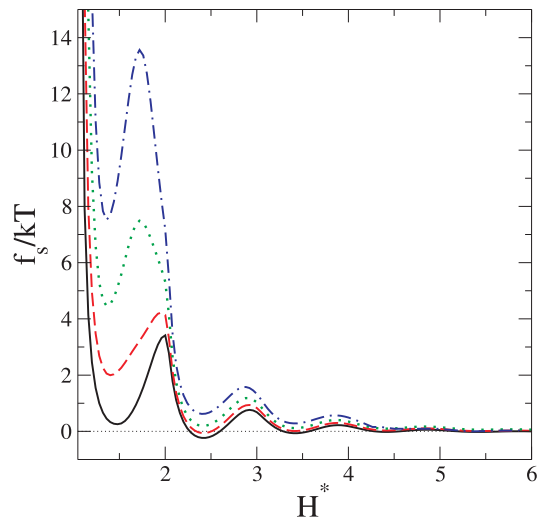}
\end{center}
\caption{(Color online) Solvation force for the systems  $5+3-$ (solid line), $6+2-$ (dashed line), $7+1-$ (dotted line)
and $8+$ (dash-dotted line). The grafting density is $R_\mathrm{C}^*=0.05$ and the bulk
electrolyte  mole fraction is $x=0.1$.
}
 \protect
\label{fig:7}
\end{figure}

The results presented above have shown that one of the important factors
effecting the behavior of the solvation force is the resultant charge of a brush.
Therefore, in figure~\ref{fig:7}, we present the results for the chains of 8 segments with the
resultant charge equal to $2e$, $4e$, $6e$ and $8e$, respectively. Now, the differences between
the solvation forces for a particular system are big. For a larger resultant charge a strong repulsion
occurs at larger wall-to-wall separations. This is obviously due to electrostatic
repulsions between the chains tethered at the opposite walls. When the resultant charge of
a chain is $2e$, we still observe effective attractions between two pore walls at $H^*\approx 2.5$ and for
$H^*\approx 3.5$.  For the resultant charge of $4e$, there appears a marginally attractive force between two walls at $z\approx 2.5$. However, for larger resultant charges, the solvation force is repulsive
for all $H^*$, although it exhibits several local minima and maxima.
When the surface grafting density increases, the local extrema become weaker, and
the curve describing the dependence of the solvation force on $H^*$ becomes smoother [cf. also figure~\ref{fig:2}~(a)].

\section{Summary}

In this work we applied the density functional theory to the study of the
solvation forces between the walls covered with tethered layers of
charged chains. We considered the chains possessing different net (resultant)
charges and investigated the cases when all the segments had the same charges,
as well as polyampholytes with zero and non-zero resultant charge per chain.
The results of calculations showed that one of the
factors that effects  the solvation force very much is the
resultant charge of the chain. However, the distribution of the charges
along the chain effects the solvation force rather little. Another
factor that quantitatively changes the solvation force is the grafting
density. On the other hand, the concentration of the solution plays
rather a minor role. Of course, the above results are valid for a
specific model studied in this work. Moreover, we considered
rather short chains and thus cannot exclude that our conclusions regarding the tethered layers built of chains involving hundreds of
segments would be different.

\section*{Acknowledgements}
The research of O.P. and J.I. leading to these results has received funding from the European Union
Seventh Framework Program ([FP7/2007--2013] under grant agreement  PIRSES~268498.
A.P. and S.S. acknowledge  support from the Ministry of Science of
Poland under the Grant No.~N~N204 151237.

\newpage
\ukrainianpart

\title{Сила  сольватації між шарами приєднаних поліелектролітів. Підхід за допомогою   функціоналу  густини}

\author{О.~Пізіо\refaddr{label1}, А.~Патрикєєв\refaddr{label2}, С.~Соколовскі\refaddr{label2}, Я.~Ільницький\refaddr{label3}}

\addresses{
\addr{label1} Iнститут хiмiї УГАМ, 04360   Мехiко,  Мексика
\addr{label2} Вiддiл моделювання фiзико-хiмiчних процесiв, унiверситет Марiї   Кюрi-Склодовської, \\
20031 Люблiн,  Польща
\addr{label3}Інститут фізики   конденсованих систем НАН України, вул. Свєнціцького, 1, 79011, Львів,   Україна
}

\makeukrtitle

\begin{abstract}
\tolerance=3000%
За допомогою варіанту   теорії функціоналу густини вивчено силу   сольватації між двома пластинами, поверхні яких модифіковано шаром приєднаних полімерних ланцюжків. Останні сформовано у вигляді з'єднаних дотичних один до   одного заряджених сферичних сегментів. Пластини занурено в розчин електроліту,   який містить катіони, аніони та молекули розчинника (які моделюються як тверді   сфери). Ми концентруємось на залежності сили сольватації та структури як   полімерних ланцюжків так і розчинника від густини приєднання ланцюжків, їх   довжини, архітектури та концентрації розчиненої   речовини.
\keywords приєднані електроліти, сила сольватації,   адсорбція, теорія функціоналу   густини

\end{abstract}

\end{document}